\documentclass[USenglish,twocolumn]{article}

\ifx\directlua\undefined\ifx\XeTeXcharclass\undefined
  \usepackage[utf8]{inputenc}                           %pdftex engine
  \else\RequirePackage[no-math]{fontspec}[2017/03/31]\fi %xetex engine
  \else\RequirePackage[no-math]{fontspec}[2017/03/31]\fi %luatex engine
\usepackage[sort&compress,square,numbers]{natbib}
\usepackage[big,online]{dgruyter}
\usepackage{color}
\usepackage{braket}

\theoremstyle{dgthm}

\theoremstyle{dgdef}

\def\rr{\mathbf{r}}

\def\MM{\boldsymbol{M}}
\def\kk{\mathbf{k}}

\def\GG{\mathbf{G}}
\def\ggg{\mathbf{g}}
\def\dd{\mathbf{d}}

\def\EE{\mathbf{E}}
\def\HH{\mathbf{H}}

\begin{document}

\articletype{Research Article}

\author*[1]{Erik P. Navarro-Barón}
\author[2]{Herbert Vinck-Posada}
\author[3]{Alejandro González-Tudela} 
\affil[1]{Universidad Nacional de Colombia, Bogotá D.C., Colombia, epnavarrob@unal.edu.co; 0000-0002-9985-3841}
\affil[2]{Universidad Nacional de Colombia, Bogotá D.C., Colombia, hvinckp@unal.edu.co}
\affil[3]{Instituto de Física Fundamental - Consejo Superior de Investigaciones Científica  (CSIC), Madrid, España, a.gonzalez.tudela@csic.es; 0000-0003-2307-6967}
\title{Directional spontaneous emission in photonic crystal slabs}
\runningtitle{Directional emission in PhC slabs}
\abstract{Spontaneous emission is one of the most fundamental out-of-equilibrium processes in which an excited quantum emitter relaxes to the ground state due to quantum fluctuations. In this process, a photon is emitted that can interact with other nearby emitters and establish quantum correlations between them, e.g., via super and subradiance effects. One way to modify these photon-mediated interactions is to alter the dipole radiation patterns of the emitter, e.g., by placing photonic crystals near them.  One recent example is the generation of strong directional emission patterns—key to enhancing super and subradiance effects—in two dimensions by employing photonic crystals with band structures characterized by linear iso-frequency contours and saddle-points. However, these studies have predominantly used oversimplified toy-models, overlooking the electromagnetic field's intricacies in actual materials, including aspects like geometrical dependencies, emitter positions, and polarization. Our study delves into the interaction between these directional emission patterns and the aforementioned variables, revealing the untapped potential to fine-tune collective quantum optical phenomena}

\keywords{Spontaneous emission, Collective phenomena; Photonic crystal slabs; quantum emitters; van Hove singularities; Light Polarization}
\journalname{Nanophotonics}
%\dedication{xxxxxxxx.}
\journalyear{2023}
\journalvolume{aop}

\maketitle

\section{Introduction} \label{sec:01}

Spontaneous emission is a paradigmatic out-of-equilibrium phenomenon that depends on both the emitter and its interaction with the surrounding media, as proposed by Purcell~\citep{Purcell1946}. It consists of a quantum system transitioning from an excited energy state to a lower energy state by emitting a photon with quantized energy equal to the difference between the two energy states. This process depends on the intrinsic properties of the emitter, such as the emitter's transition frequency and dipole moment, as well as the surrounding electromagnetic environment, which can be modified by placing nearby materials. This opens the exciting possibility of tuning such fundamental phenomena by positioning the emitter close to macroscopic materials, like cavities~\citep{haroche89a}, placing nearby emitters~\cite{lehmberg70a,lehmberg70b}, or combinations of both.

One of the most paradigmatic examples of modifying individual and collective spontaneous emission of quantum emitters are photonic crystals (PhC)~\cite{joannopoulos97a,lodahl15a, Chang2018}, which are now routinely interfaced with many different types of emitters~\cite{goban13a,thompson13a,hood16a, Beguin2020a, Zhou2023,evans18a, Samutpraphoot2020, Dordevic2021, MenonAnImaging, Tiranov2023}. In these systems, the propagation of light becomes modulated by the periodic variation of the refractive index, resulting in non-trivial band structures, e.g., featuring photonic band-gaps~\cite{Yablonovitch1987}, Dirac-like energy dispersions~\cite{bravo12a} or saddle-points~\citep{Mekis1999, Witzens2002, Lin2013}. These can substantially modify the individual and collective radiative patterns of emitters placed nearby. For example, when the emitter's optical transition lies within photonic band-gaps, spontaneous emission is largely suppressed, leading to the formation of atom-photon bound states~\cite{bykov75a,kurizki90a,john90a,john94a}. These states can induce perfect coherent exchanges between emitters with different shapes depending on the lattice geometry~\cite{douglas15a, Gonzalez-Tudela2015b, Gonzalez-Tudela2019a, Perczel2018a, Navarro-Baron2021, Bello2019a, Garcia-Elcano2020}. In the opposite regime, when the emitter lies within the band, spontaneous emission occurs but with renormalized lifetimes and radiation patterns. A particularly intriguing situation is the spontaneous emission modification that occurs in two-dimensional photonic crystals due to the presence of saddle-points and straight iso-frequencies in the band structures~\citep{Mekis1999, Witzens2002}. While it has long been known that such energy dispersions can lead to unconventional light propagation, such as supercollimation~\citep{Mekis1999, Witzens2002}, it has been recently pointed out that when emitters are resonant to these frequency regions, spontaneous emission can be accelerated to the point where it becomes non-perturbative~\cite{Gonzalez-Tudela2017a}, and their individual radiation patterns become highly directional~\citep{Gonzalez-Tudela2017, Gonzalez-Tudela2017a, Galve2017, Galve2018, Yu2019} leading to strong super and subradiant effects when several emitters are present~\cite{Gonzalez-Tudela2017, Gonzalez-Tudela2017a}. However, most of these predictions were made for simplified coupled-mode descriptions~\citep{Gonzalez-Tudela2017, Gonzalez-Tudela2017a, Galve2017, Galve2018}, in which the complexity of the electromagnetic field, such as its geometrical dependence or polarization, were neglected, or for particular emitter positions~\cite{Witzens2002, Lin2013, Yu2019}. The interplay of these factors with such unconventional individual and collective phenomena remains an open question.

In this work, we undertake the challenge of studying the interplay of such directional emission with the emitter's 
dipole polarization and the emitter's position in a particular two-dimensional PhC slab, i.e., the honeycomb structure studied in Refs.~\cite{Perczel2018a, Navarro-Baron2021}. To achieve this, we theoretically investigate the electromagnetic wave propagation of dipole emitters with different positions and polarizations using the Guided Mode Expansion (GME)~\citep{Andreani2006d} and the finite difference time domain (FDTD)~\citep{Taflove2005} algorithms. From this study, we observe the expected directional emission into six rays predicted by toy-models~\cite{Gonzalez-Tudela2018} but also note that some of these directions can be enhanced or suppressed depending on the emitter's polarization and position. The manuscript is organized as follows: In Sec.~\ref{sec:02}, we introduce the PhC structure we are considering and the theoretical methods and foundations used in this work. Sec.~\ref{sec:03} discusses the relationship between the classical Green's function and the individual and collective spontaneous emission phenomena. In Sec.~\ref{sec:04}, we describe the emission of a dipole emitter at the center of the unit cell with circular polarization. Sec.~\ref{sec:05} examines different emitter positions and dipole moment polarizations to analyze their interaction with directional emission. Finally, we summarize our findings and outline potential applications in Sec.~\ref{sec:06}.

\section{Photonic structure}\label{sec:02}

\begin{figure}[tb!]
    \includegraphics[scale=0.35]{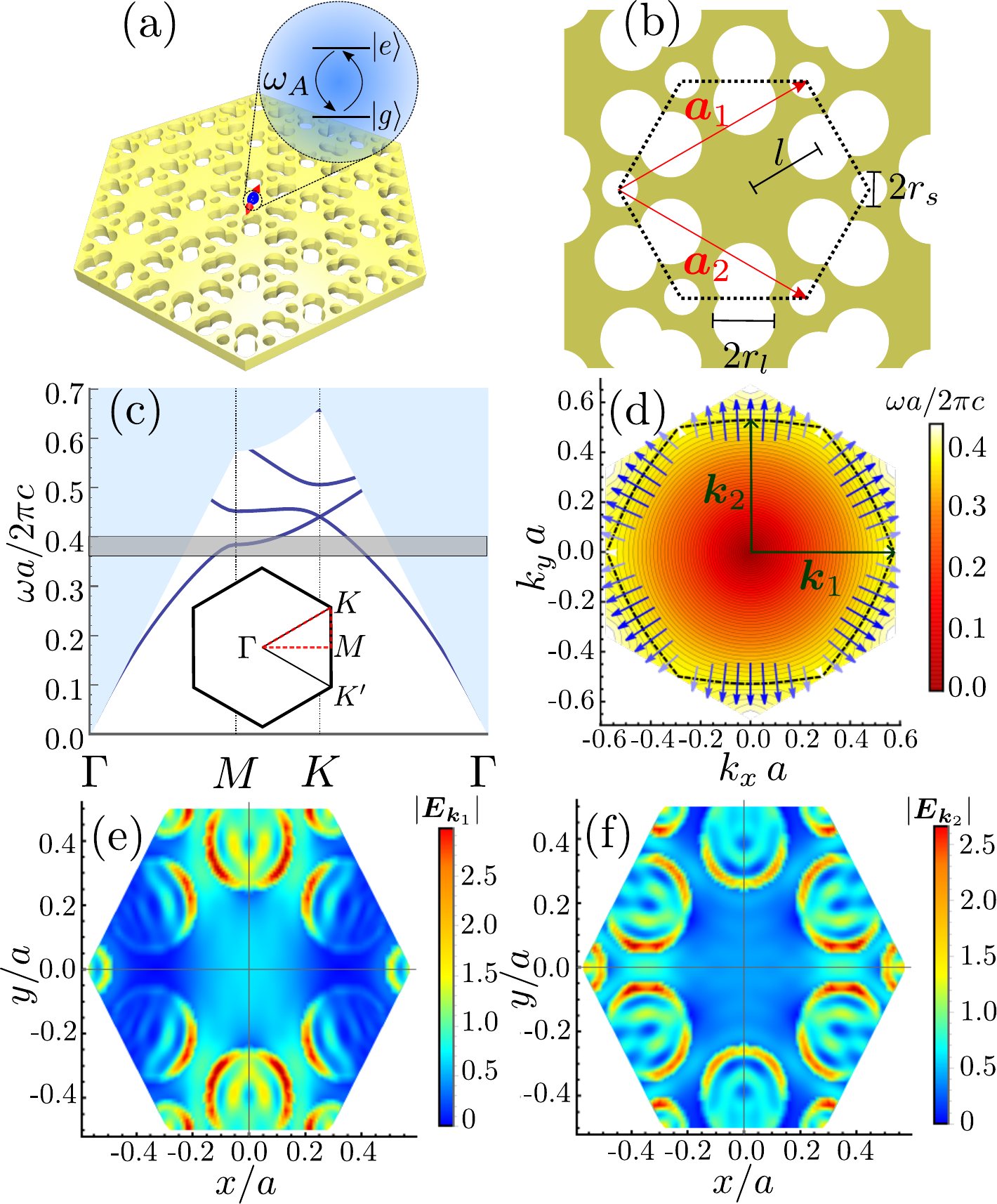}
    \caption{Photonic crystal structure. (a) Schematic representation of an atom-like emitter embedded in a photonic crystal slab of a hexagonal lattice with periodic parameter $a$  and thickness $d=0.25a$, the frequency emission source is $\omega_A$, this photonic structure was proposed in previous work~\citep{Perczel2018a}. (b) Top view of the photonic crystal slab, the yellow region represents a dielectric media with $\varepsilon_{GaP}= 10.5625$, and the white regions show air holes. The unit cell has six smaller holes of radii $r_s=0.0833 a$ and six bigger holes of radii $r_l=0.15 a$ with center at a distance $l=0.4a$ of the unit cell center. (c) Band diagram of the PhC slab in the inverse Brillouin Zone edges (red dashed line in the inset), the gray shaded region shows the frequency region of the van Hove singularity about $\omega_M=0.3849 \left(\frac{2\pi c}{a}\right)$.(e) and (f) Electric field amplitude for the first band mode for $k_1$(e) and $k_2$(f), the vectors $k_1$ and $k_2$ in the irreducible Brillouin Zone are showed as green arrows in (d).}
    \label{fig:01system}
\end{figure}

We consider the photonic-crystal (PhC) slab depicted in Fig.~\ref{fig:01system}, introduced for the first time in Ref.~\cite{Perczel2018a} as a way of frequency-isolating Dirac points. The slab consists of a hexagonal lattice of GaP ($\varepsilon_{\mathrm{GaP}}=10.5625$) with a thickness $d$ and primitive vectors $\mathbf{a}_{1/2}=(\sqrt{3}/2,\pm1/2)a$, where $a$ is the lattice constant. The slab has six bigger and six smaller air holes of radii $r_l$ and $r_s$, respectively, as shown in Fig.~\ref{fig:01system}(b). The bigger holes have a distance to the unit cell center, denoted by $l$.

To compute the eigenfrequencies and eigenmodes of this PhC structure, we use the GME proposed in Ref.~\citep{Navarro-Baron2021} for this structure. The GME method uses the solutions of an effective homogeneous slab denoted by $\boldsymbol{h}_{\ggg,\mu}(\rr,z)$ to expand the modes of the PhC slab. Here, we use $\rr$ to denote the position in the plane of periodicity ($xy$ plane) and put explicitly the $z$ coordinate. In principle, the homogeneous slab has two types of solutions: radiative modes oscillating outside the slabs and guided modes that decay exponentially outside the slabs. Initially, the method considers only the guided modes that are characterized by $\boldsymbol{h}_{\ggg,\mu}(\rr,z)= e^{i\ggg\cdot\rr}\mathbf{h}_{\ggg,\mu}(\rr,z)$~\citep{Andreani2006d, Navarro-Baron2021}, where the vector $\ggg=\kk+\GG$ is the wave vector in the $xy$ plane, with $\GG$ a reciprocal lattice vector that contains information of the system periodicity and $\kk$ a wave vector in the first Brillouin zone; and $\mu$ is an index for the quantization of the guided modes. By using these modes, the magnetic field for the PhC slabs is expanded in the frequency domain (harmonic fields) as:
\begin{equation}
    \mathbf{H}_{\kk,n}(\rr,z)=\sum_{\GG,\mu} c_n(\kk+\GG,\mu) \boldsymbol{h}_{\kk+\GG,\mu}(\rr,z)\, ,
    \label{Ec:1-GME MagneticField}
\end{equation}
where $c_n(\kk+\GG,\mu)$ are the coefficients of expansion and $n$ denotes the different modes of the PhC slab that appear with frequency $\omega_{\kk,n}$. A matrix eigenvalue problem is obtained by putting these fields in the time-independent wave equation for the magnetic field 
\begin{equation}
    \sum_{\ggg,\nu}\mathcal{H}_{\ggg,\mu;\ggg',\nu} c_n(\ggg,\nu)=\left(\frac{\omega_{\kk,n}}{c}\right)^{2}c_n(\ggg,\mu)\, ,
    \label{Ec:4-GME-MatrixEigenProblem}
\end{equation}
where the sum over $\ggg=\kk+\GG$ is equivalent to the sum over the lattice vector $\GG$ because $\kk$ is fixed for each eigenvalue calculation. The matrix elements $ \mathcal{H}_{\ggg,\mu;\ggg',\nu}$ read
\begin{equation}
    \mathcal{H}_{\ggg,\mu;\ggg',\nu}=\int_{V_c}\frac{1}{\varepsilon(\rr)}(\nabla\times\boldsymbol{h}^{*}_{\ggg,\mu}(\rr))\cdot(\nabla\times\boldsymbol{h}_{\ggg',\nu}(\rr))d\rr dz\, ,
\end{equation}
where $V_c$ is the integration volume defined by the unit cell in the $xy$-plane and $(-\infty,\infty)$ in the $z$-direction; details about the analytic expression for matrix elements can be found in Ref.~\citep{Andreani2006d}. In principle, equation \eqref{Ec:4-GME-MatrixEigenProblem} considers infinite guided modes in the expansion of fields, so the eigenproblem is infinite. However, the guided modes basis must be truncated to a finite size to compute it numerically. Solving the finite eigenproblem, we obtain the eigenvalues $\frac{\omega_{\kk,n}^2}{c^2}$ and the eigenvectors containing the coefficients $c_n(\ggg,\nu)$ that expand the magnetic field mode $\mathbf{H}_{\kk,n}(\rr,z)$ in the truncated basis. The band diagram of our photonic structure is displayed in Fig.~\ref{fig:01system}(c) for the TE-like (even) modes. We are interested in describing the emission of the dipole source in the middle of the slab ($z=0$) with components in the $xy$ plane, so the TM-like (odd) modes are neglected because they have only $z$-component at $z=0$. Here, the first band has a frequency region near the symmetry point $\MM=\frac{2\pi}{3a}(\sqrt{3},0)$ that presents a saddle-point, which will result in an associated Van-Hove singularity (VHS).  To analyze in detail the directional emission in this frequency region, it is necessary to consider the contribution of all the $\kk$ vectors in the first Brillouin, which is why in Fig.~\ref{fig:01system}(d), we plot a density plot of the first band. The iso-frequency curve at $\omega_{\MM}$ is displayed by the black-dashed line showing a quasi-straight hexagonal curve, which suggests that the contribution of this $\kk$ vector propagate in six direction perpendicular to each side of the hexagon. Additionally, for all the $\kk$ vectors in this curve, we calculate the group velocity through the following relation~\citep{Joannopoulos2011}:
\begin{align}
\mathbf{v}_{g}= \frac{\int_{\partial V_c} \mathbf{S} d^3\rr }{U_{\EE}+U_{\HH}}\,,
\label{Ec:04-groupvelocity}
\end{align}
where $\mathbf{S}$ corresponds to the Poynting vector, and $U_{\EE(\HH)}$ denotes electric(magnetic) energy density integrated into the volume $V_c$. To compute the Poynting vector, we use the magnetic field previously calculated using the GME method. The group velocity for the $\kk$ vectors in the iso-frequency curve point almost preferentially in six directions, each one perpendicular to a side of the iso-frequency, and the maximum contribution of the group velocity is at specific $\kk$ vectors in these specific directions, that are angles of $n\pi/3+\pi/6$ rad, with $n=0,1, \dots, 5$. These results explain the self-collimation phenomena that cause the directional emission~\citep{Witzens2002, Lin2013}. Finally, in Fig.~\ref{fig:01system}(e,f), we present the electric field distribution for two $\kk$ vectors in the iso-frequency curve that are represented with green arrows in the panel (d) of the same figure. The field distribution shows that for every single $\kk$, the maximum field in the unit cell changes. This ultimately means that the coupling of the emitter with the PhC structure will strongly depend on which position in the unit cell the emitter has. This is what we will explore in the next sections.

\section{Classical Green's function and quantum spontaneous emission}\label{sec:03}

To calculate the effect of the emitter positions and the field polarization in the directional spontaneous emission near PhC, we use the macroscopic QED formalism~\cite{scheel08a}. This formalism gives an effective description for the quantum emitter dynamics after tracing the photonic degrees of freedom of the PhC. Assuming we have $N$ identical emitters with transition frequency $\omega_A$, the description is given in terms of a Born-Markov master equation which reads:
\begin{equation}
    \frac{d\rho(t)}{dt}=-\frac{i}{\hbar}[H_0+H_\mathrm{eff},\rho]+\mathcal{L}_\mathrm{eff}(\rho)\,, \label{eq:meq}
\end{equation}
where $\rho$ is the density matrix describing the emitter degrees of freedom, $H_0= \sum_{j}\hbar\omega_A\sigma^j_{ee}$ corresponds to the independent emitters' Hamiltonian, and $\sigma^j_{\alpha\beta}=\ket{\alpha}_j\bra{\beta}$ is the notation introduced for the emitter operators. Besides, as a result of the interaction with the photons, the emitters' dynamics contain two additional terms, $H_\mathrm{eff}$ that represents the coherent part of the photon-mediated interactions, which read:
\begin{align}
H_{\mathrm{eff}}=\hbar\sum_{i,j} J_{ij}\sigma_{eg}^i\sigma_{ge}^j\,.\label{eq:Hij}
\end{align}
and, $\mathcal{L}_\mathrm{eff}(\rho)$ that describes the incoherent dynamics induced by the photonic bath that reads:
\begin{align}
\mathcal{L}_\mathrm{eff}(\rho)=\sum_{ij}\frac{\Gamma_{ij}}{2}\left(2\sigma_{ge}^i\rho\sigma_{eg}^j-\sigma_{eg}^i\sigma_{ge}^j\rho-\rho\sigma_{eg}^i\sigma_{ge}^j\right)\,.\label{eq:Gammaij}
\end{align}

\begin{figure}[tb!]
    \centering
    \includegraphics[scale=0.8]{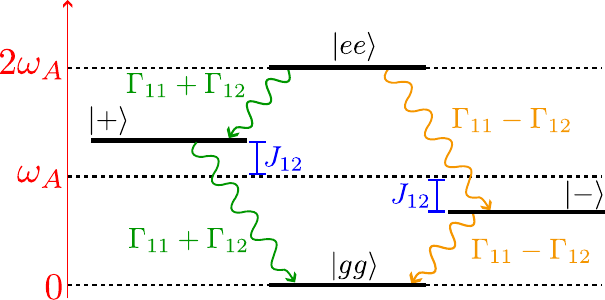}
    \caption{Schematic representations of the states and the renormalization produced in the energies and emission rates by the mediated interactions.}
    \label{fig:02-quantum-mediated-interactions}
\end{figure}

The key point of this formalism is that both the coherent and incoherent photon-mediated interactions, $J_{i,j}$ and $\Gamma_{i,j}$, respectively, are expressed in terms of the classical Green's function as follows:
\begin{align}
J_{ij}&=-\frac{\mu_0\omega_A^2}{\hbar}\dd^{*}_i\cdot\mathrm{Re}[\GG(\rr_i,\rr_j,\omega_A)]\cdot\dd_j\,, \\
\Gamma_{ij}&=\frac{2\mu_0\omega_A^2}{\hbar}\dd^{*}_i\cdot \mathrm{Im}[\GG(\rr_i,\rr_j,\omega_A)]\cdot\dd_j\,,
\end{align}
with $\dd_j$ being the dipole moment of the $j$-th atom. Here, $\Gamma_{i,j}$ is known as the cooperative or collective decay rate for $i\neq j$, and corresponds to the spontaneous emission rate of a single emitter for $i=j$. When several emitters are present, both the individual and collective decay terms strongly renormalize their spontaneous emission. To evidence it, we plot schematically the effect of $\Gamma_{i,j}$ and $J_{i,j}$ in the Hilbert space of two two-level emitters composed by the states: $\ket{gg}= \ket{g}\otimes \ket{g}$; $\ket{\pm}=(\ket{e}\otimes\ket{g}\pm\ket{g}\otimes\ket{e})/\sqrt{2}$; $\ket{ee}=\ket{e}\otimes\ket{e}$, shown in Figure~\ref{fig:02-quantum-mediated-interactions}. There, we observe how $J_{ij}$ shifts the energies of the $\ket{\pm}$ states, whereas $\Gamma_{ij}$ renormalize their lifetime to be $\Gamma_{ii}\pm\Gamma_{ij}$. 

Summing up, calculating the real and imaginary parts of classical Green's function, we indirectly have access to the collective dynamics of emitters. This is why, in the next sections, we focus precisely on calculating the classical Green's tensor defined by the equation:
\begin{equation}
    \nabla \times  \nabla \times \GG(\rr,\rr',\omega)-\frac{\omega^2}{c^2}\varepsilon(\rr,\omega)\GG(\rr,\rr',\omega)=\mathbb{I} \delta(\rr-\rr')\,,
\end{equation}
where, $\GG(\rr,\rr',\omega)$ denotes the Green's tensor with components $G_{\alpha,\beta}(\rr,\rr',\omega)$. Its classical meaning is that a point source oscillating with frequency $\omega$ and dipole moment $\mathbf{d}$ at the position $\rr'$ produces an electric field at the position $\rr$ determined by the following expression:
\begin{equation}
    \EE(\rr,\omega)=\mu_0 \omega^2 \GG(\rr,\rr',\omega)\cdot \mathbf{d},
\end{equation}

To find Green's function, we use an FDTD algorithm that numerically calculates the field in the time domain produced by a dipole source. Finally, to obtain Green's tensor in the frequency domain, we perform a fast Fourier transform considering a time window long enough not to include the effects of turning the source on/off. By simulating the field produced by a dipole with only $x$-component, we can calculate the Green tensor components $G_{xy}$ and $G_{xx}$ and do the same for a dipole with $y$-component to find $G_{yy}$ and $G_{yx}$. In what follows, we are interested in putting the source in the middle of the slab where the $z$-component of the TE-like modes is zero, so $G_{z\beta}=G_{\alpha z}=0$.

\section{Directional spontaneous emission}\label{sec:04}

As mentioned in the introduction, recent works in the literature with toy-model honeycomb geometries~\cite{Gonzalez-Tudela2018} predict a highly directional radiation pattern when the emitter is tuned to frequency regions with straight iso-frequencies and saddle-points. In this section, we intend to explore what happens in the PhC slab structure described in Section~\ref{sec:02}. For that, we use Ansys Lumerical - FDTD software to numerically simulate a PhC slab in a finite hexagonal region with 19 unit cells from side to side. After that, we consider a dielectric homogeneous slab with permittivity $\varepsilon_{GaP}$. For concreteness, we consider a lattice parameter $a=0.28044\times10^{-6}$ m, corresponding with a frequency $f_{\MM}=\omega_{\MM}/2\pi=399,595$ THz. The rest of the geometrical parameters are fixed in terms of $a$ equal to the ones described in Fig.~\ref{fig:01system}.

\begin{figure}[tb!]
    \centering
    \includegraphics[scale=0.38]{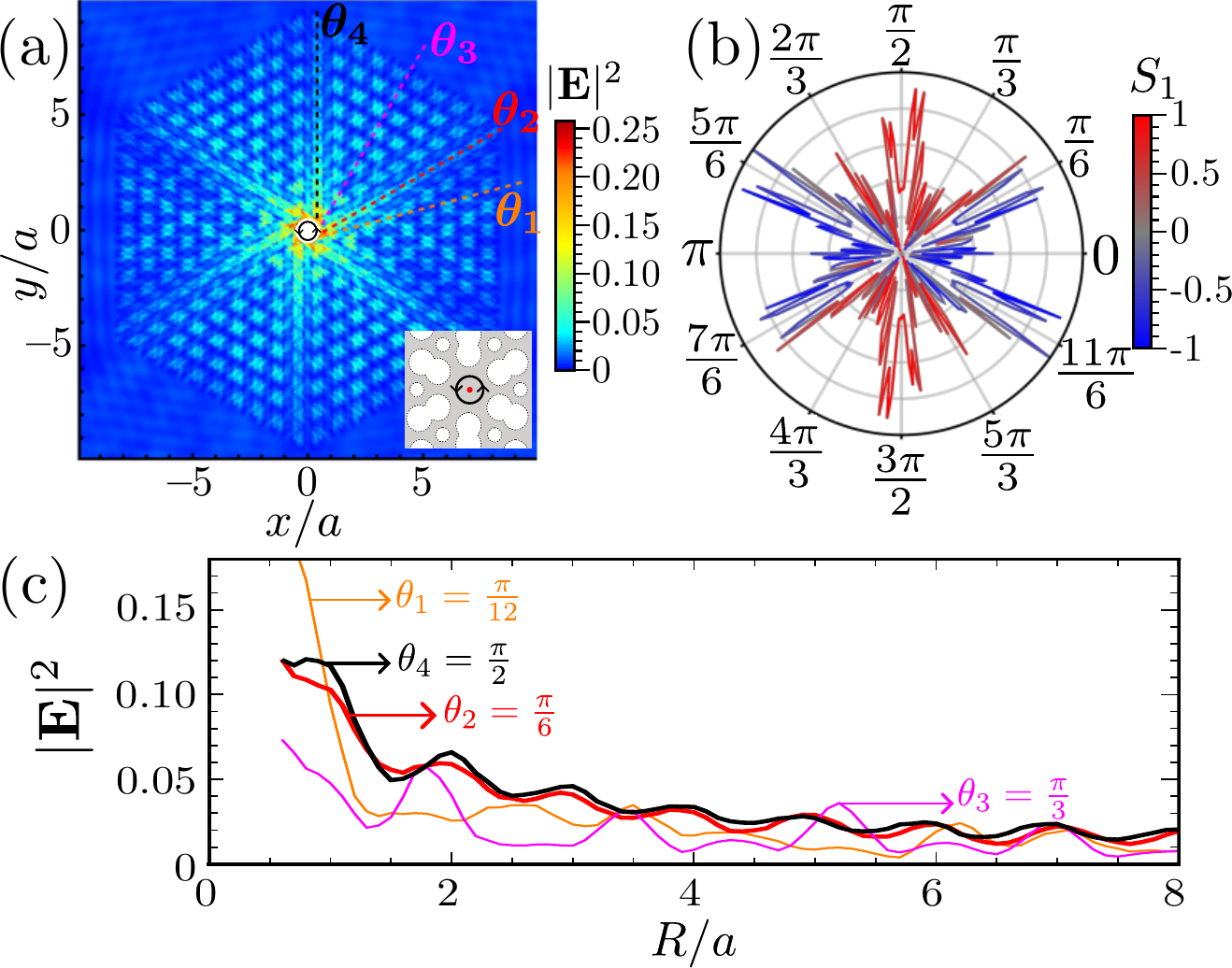}
    \caption{Directional emission in six paths. (a) A density plot of the electric field intensity (V\textsuperscript{2}/m\textsuperscript{2}) of a point dipole source with $\sigma_+$ polarization at the center of the unit cell as depicted in the inset. The dashed lines show four directions to analyze field propagation, as shown in (c). (b) The polar plot of the electric field intensity at a fixed distance of dipole $r=4a$, the radius of the polar plot represents the field intensity, and the color represents the Stokes' parameter $S_1$, indicating the linear polarization of the field. (c) Electric field intensity as a function of distance to the source in four different directions ($\theta_1=\pi/12$, $\theta_2=\pi/6$, $\theta_3=\pi/3$ and $\theta_4=\pi/2$ displayed en (a)). It is noticed that the field intensity for the $\theta_2$ and $\theta_4$ is bigger than the other in almost all distances $R$.}
    \label{fig:02dir-emission}
\end{figure}

We start considering a dipole source at the center of the unit cell with a frequency $f_{\MM}$ and a left-handed circular polarized ($\sigma_+$) dipole moment, i.e., $\mathbf{d}=\mathsf{d}\hat\sigma_+$, where $\hat\sigma_\pm=\mp (\hat x\pm i \hat y )/\sqrt{2}$. In Fig.~\ref{fig:02dir-emission}(a), we plot the corresponding electric field intensity radiated into the structure, showing that in this case, the emitter radiates into the structure along in six directions at $\pm \pi/6$ rad, $\pm \pi/2$ rad, and $\pm 5\pi/6$ rad. Thus, for this emitter's position and polarization, the behavior is qualitatively similar to the one predicted by toy-models~\cite{Gonzalez-Tudela2018}. 

To make a more detailed analysis of the radiation, in Fig.~\ref{fig:02dir-emission}(b), we make a polar plot of the electrical field intensity at a fixed distance $r=4a$ from the source in terms of the angular coordinate $\theta$. Like this, it is more clear that field intensity, represented by the radial coordinate, is maximal slightly shifted from the $n\pi/3+\pi/6$ rad. The underlying reason is that the geometrical structure of the unit cell favors the field to accumulate near the air holes. Apart from that intensity information, we also include a color map in the polar plot, which represents the $S_1$ Stokes parameter defined as:
\begin{equation}
    S_1= \frac{E_x^2-E_y^2}{|\EE|^2},
\end{equation}
where, $E_{x(y)}$ represents the $x(y)$-component of the electric field. With this definition, $S_1=1$ (red color) indicates horizontal (H) polarization, while $S_1=-1$ (blue color) indicates the vertical (V) one~\citep{Lang2015, Sauer2020}.  With this color code, one can immediately notice a strong relation between the direction of propagation and the polarization, reminiscent of what happens in one-dimensional chiral quantum optical systems~\cite{lodahl17a}, where certain propagation directions are locked to certain polarizations. For example, along the vertical propagation around $\pm\pi/2$ rad, the field is clearly H-polarized, while in oblique directions, the V polarization dominates. These results suggest that selecting the emission direction using different linear polarization should be possible, as we explore in the next section. 

Finally, to ensure that the peaks in the polar plot are not an artifact of the distance chosen ($4a$), we have analyzed the field intensity as a function of the distance for the four directions depicted with dashed lines in Fig.~\ref{fig:02dir-emission}(a). The field intensity in these directions is plotted in Fig.~\ref{fig:02dir-emission}(c), showing that directions at $\pi/6$ rad and $\pi/2$ rad dominate over the other one at $\pi/12$ and $\pi/3$ in almost all distances. 

\section{Interplay of directional emission with field polarization and emitter position}\label{sec:05}

From the results obtained in the previous section, it became clear that there is a distinct interplay between the direction of propagation of the field and its polarization. In this section, we aim to explore how one can use this degree of freedom, or others like the emitter's position, to tune the directional emission patterns. In particular, we will first study the dependence of linearly polarized dipoles at the center of the unit cell, and then consider the effect of the emitters' positions by moving them to locations outside of the unit cell center.

\subsection{Linear polarization}

\begin{figure}[tb!]
    \centering
    \includegraphics[scale=0.38]{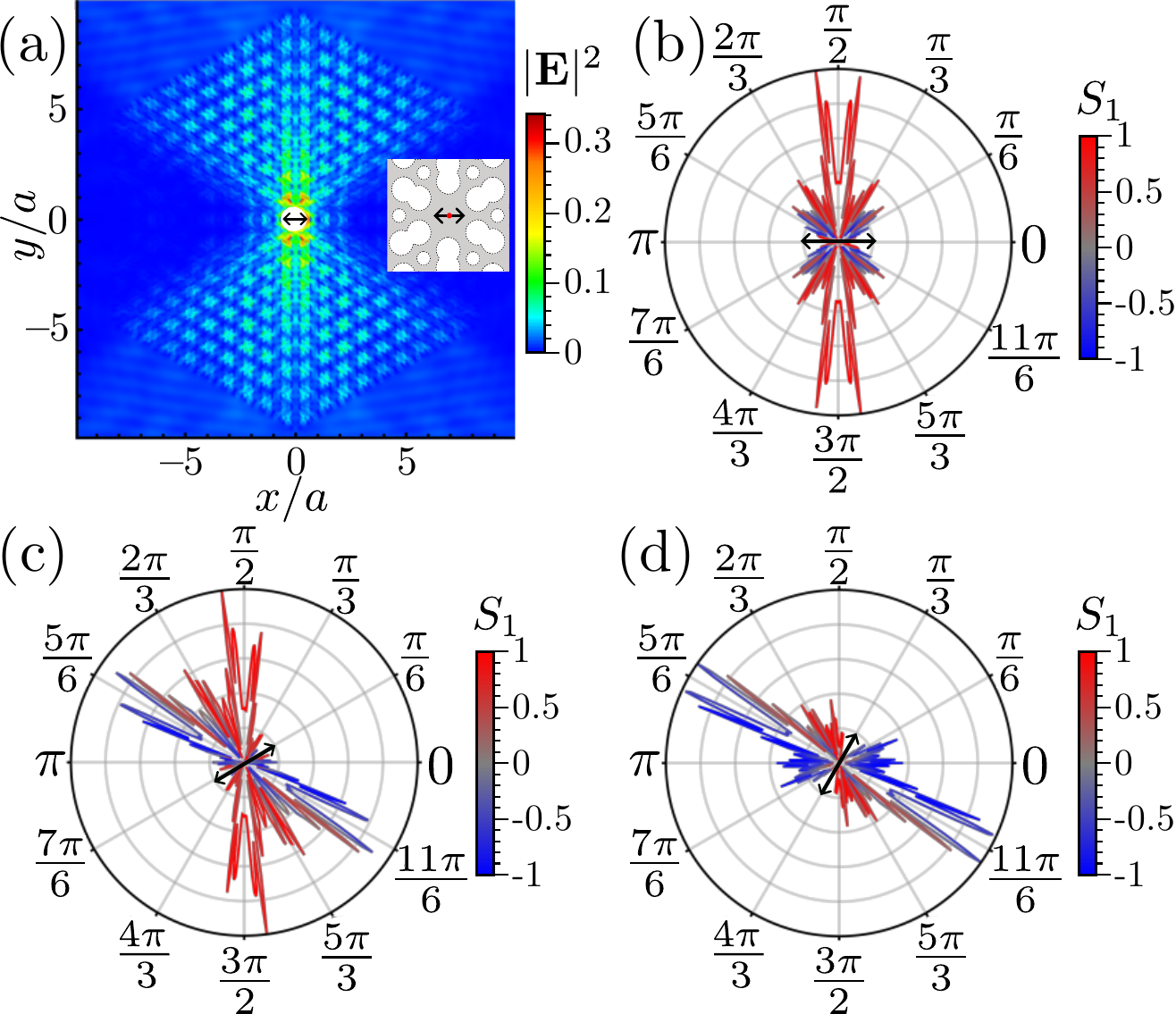}
    \caption{Choosing paths with linear polarization. (a) Intensity of the electric field from a linear polarized dipole source with H polarization at the center of the unit cell, as depicted in the inset. (b-d) Polar plot of the electric field intensity at a distance of $4a$ from the source with dipole orientation at $0$ rad [(b)], $\pi/6$ rad [(c)], and $\pi/3$ rad [(d)] as indicates the black arrow in each panel; the radial coordinate display the intensity at each angular direction. The color scale of the line represents the $S_1$ parameter, indicating the dominant linear polarization of the field, H polarization (red color), V polarization (blue color), and similar contribution (gray color).}
    \label{fig:03linear-polar}
\end{figure}

Let us here consider a linear dipole source at the center of the unit cell. As discussed above, the paths in the vertical direction ($\theta=\pi/2$ rad) have H polarization, so we start with a linear dipole in this polarization at the center of the unit cell, as shown in the inset of Fig.~\ref{fig:03linear-polar}. In this case, the dipole emitted light preferentially in the vertical direction, as the intensity profile shown in Fig~\ref{fig:03linear-polar}(a). This figure shows the maximum field intensity at angles around $\pm\pi/2$ and no fields at the other four main paths ($\pm \pi/6$ and $\pm 5\pi/6$ rad). To observe more clearly the directional emission, we appeal to the polar plot previously explained in Sec.~\ref{sec:04}. In the case of H polarization, the polar plot in Fig.~\ref{fig:03linear-polar}(b) shows only peaks at $\pm \pi/2$ rad, which means that only one of the three main directions was selected by this linear polarization, while the other directions are turned off.

Additionally, we analyze other dipole moment orientations, shown in panels (c-d) of Fig.~\ref{fig:03linear-polar}. In panel (c), we consider a dipole moment $\mathbf{d}= \sqrt{3}/2 \hat{x}+1/2 \hat{y}$, which excites the paths at $-\pi/6$, $\pm \pi/2$ and $5\pi/6$ rad. In this oblique case, two directions of propagation are activated because the perpendicular direction of the dipole orientation corresponds to $4\pi/6$ rad, which is in the middle of the two excited directions. This is because, in this case, the dipole couples with the same amplitude to the modes of the PhC slab that propagate in these two main directions and does not couple to the modes propagating in the third direction ($\pi/6$ rad). Similar to this case, one can consider any arbitrary orientation of the dipole to excite two of the three directions, except when considering orientations of $0$, $\pm \pi/3$, and $\pm 2\pi/3$ rad.  Finally, the particular cases of dipole orientations $0$, $\pm \pi/3$, and $\pm 2\pi/3$ rad select only one of the three main directions corresponding with the perpendicular direction of the dipole orientations, as shown in Fig.~\ref{fig:03linear-polar}(d) for a dipole orientation of $\pi/3$ rad. In the Supplementary Material, we provide an animated video displaying the field intensity distribution produced by different dipole orientations (see Supplementary Material, Video 1), showing continuously how the radiation patterns are modified as one moves the polarization of the dipole continuously.

Let us now discuss the interplay between the field polarization and the three directions in more detail. In the vertical directions, we have H polarization equal to the dipole orientation that excites this direction. However, in the oblique directions, the polarization becomes a linear combination of both H and V polarization equal to the combination of dipole orientation that activated these oblique directions at $\pi/6$ and $5\pi/6$ rad, where the dipole orientation and field polarization are defined by $\mathbf{d}=\pm \sqrt{3}/2\hat{x} \pm 1/2 \hat{y}$. This possibility of choosing the direction by using linear polarized dipoles is allowed by the fact that the electric field at the unit cell center always has a non-zero value for all modes in the iso-frequency, and thus, the dipole source can always couple to all momenta irrespective of the polarization. As we will see in the next section, this is not true for all positions of the unit cell.

\subsection{Linear polarization and emitter out of center}

\begin{figure}[tb!]
    \centering
    \includegraphics[scale=0.37]{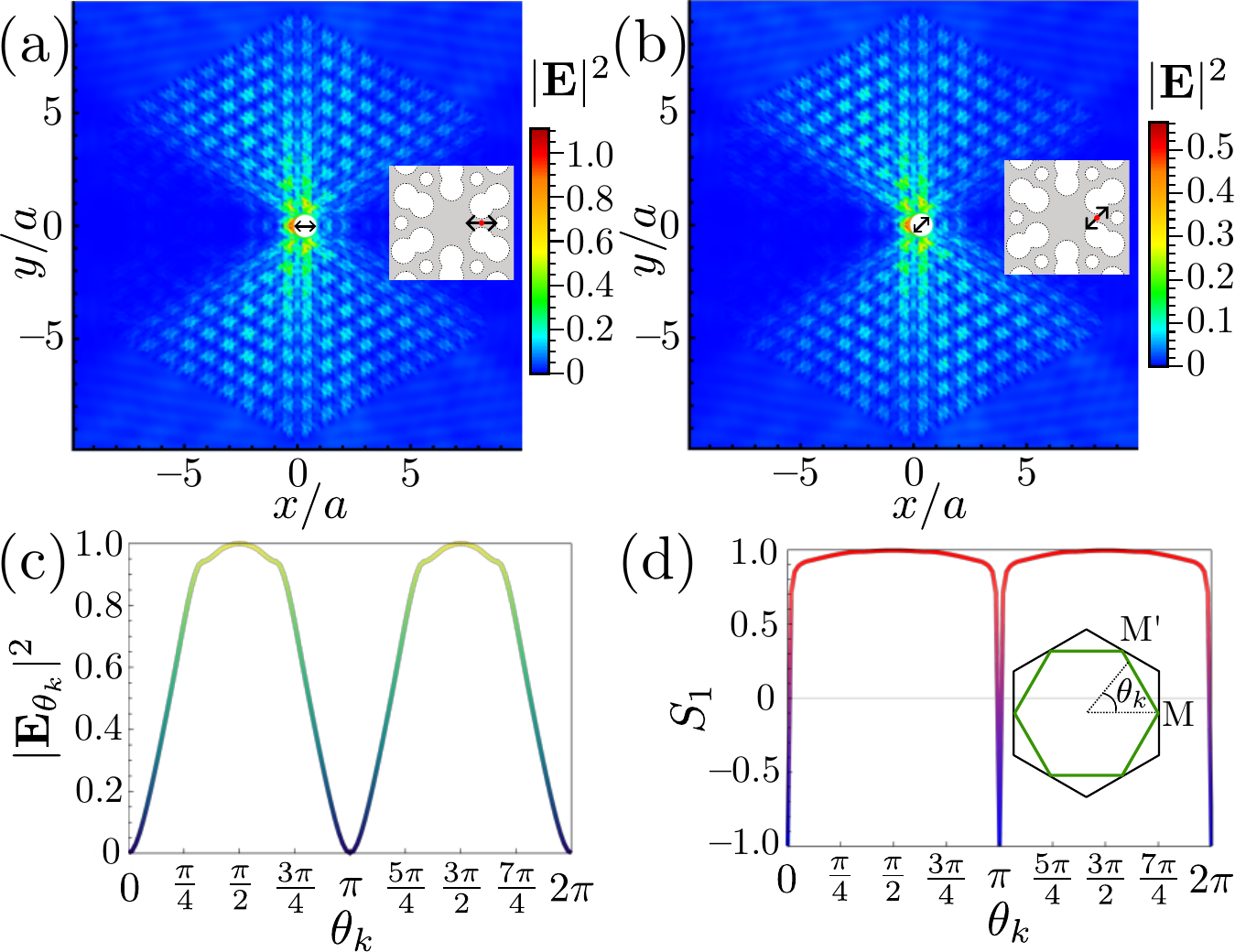}
    \caption{Linear polarized dipoles outside the center at $\rr=(0,0.35a)$. (a) Electric field intensity of an H-polarized dipole in the whole PhC slab. (b) Electric field intensity in the whole PhC slab produced by a dipole moment $\mathbf{d}=(\hat{x}+\hat{y})/\sqrt{2}$. To compare (a) and (b), one has to take into account the color scale in each panel. (c) Electric field intensity at $\rr=(0,0.35a)$ for all the modes inside the iso-frequency curve. The wave vector of each mode is represented by its angular coordinate $\theta_k$, as depicted in the inset of (d). (d) Stokes parameter $S_1$ at the position $\rr=$ for all the modes in the iso-frequency curve.}
    \label{fig:04filtered-linear}
\end{figure}

When the emitter position is the unit cell center, the PhC modes at all momenta contribute to the propagation, including all possible linear polarization contributions. However, there are other positions outside the center where some modes have a non-zero electric field while others do not have an electric field, see panels (e-f) of Fig.~\ref{fig:01system}. There, we observe modes for two wave vectors in the iso-frequency curve, one pointing in the horizontal direction (panel e) and the other in the vertical (panel f). The latter has a null electric field in the dielectric region between the two nearest bigger air holes, for example, around $\rr=(0, 0.35 a)$. In contrast, the former has a significant field contribution in this region. This difference leads to substantial differences in the control of the directional emission. To evidence that, let us consider a linear polarized dipole at position $\rr=~(0, 0.35 a)$. In this situation, the intensity profile of the electric field displays a vertical wave propagation as shown for an H-polarized dipole in Fig.~\ref{fig:04filtered-linear}(a), similar to what happens at the center of the unit cell. However, if the dipole orientation changes to an oblique angle, i.e., $\mathbf{d}=(\hat{x}+\hat{y})/\sqrt{2}$, the intensity profile is still vertical, as shown in Fig.~\ref{fig:04filtered-linear}(b). Let us, however, note the maximum intensities in the color scale, where we can observe that the intensity decreases by approximately a factor of two from the H-polarized dipole to the oblique one. Similar results for other orientations and circularly polarized dipoles are found, as shown in the videos we provide in the Supplementary Material (see Videos 2-3). Additionally, we must mention that the vertical propagation has a well-defined H polarization independent of the dipole moment.

To understand the origin of this directionality that is independent of the polarization, we analyze the contribution of each mode from the iso-frequency curve to the field propagation by use of the field intensity [Fig.~\ref{fig:04filtered-linear}(c)] and Stokes parameter [Fig.~\ref{fig:04filtered-linear}(d)] of each mode at this position $\rr=(0,0.35a)$. First, we discuss it in terms of the intensity that gives us information about the coupling of the dipole with each mode. We find that the modes with more intensity correspond to the wave vectors near the vertical directions, which explay why they dominate the radiative decay of an emitter in this position. This explains the directional emission, but it does not explain neither its polarization nature nor the intensity decrease from panel (a) to (b) of Fig.~\ref{fig:04filtered-linear}. To understand the invariant polarization of the emitted field, we examine the $S_1$ parameter at $\rr=(0,0.35a)$ for all the modes in the iso-frequency, as shown in  Fig.~\ref{fig:04filtered-linear}(d). Here, we observe that almost all the modes in the iso-frequency curve have only H polarization, which means that only the horizontal component of the dipole couples to the PhC structure and couples with half intensity due to the smaller overlap with the predominant polarization of the modes.

\subsection{Circular polarization and emitter out of center}

\begin{figure}[tb!]
    \centering
    \includegraphics[scale=0.37]{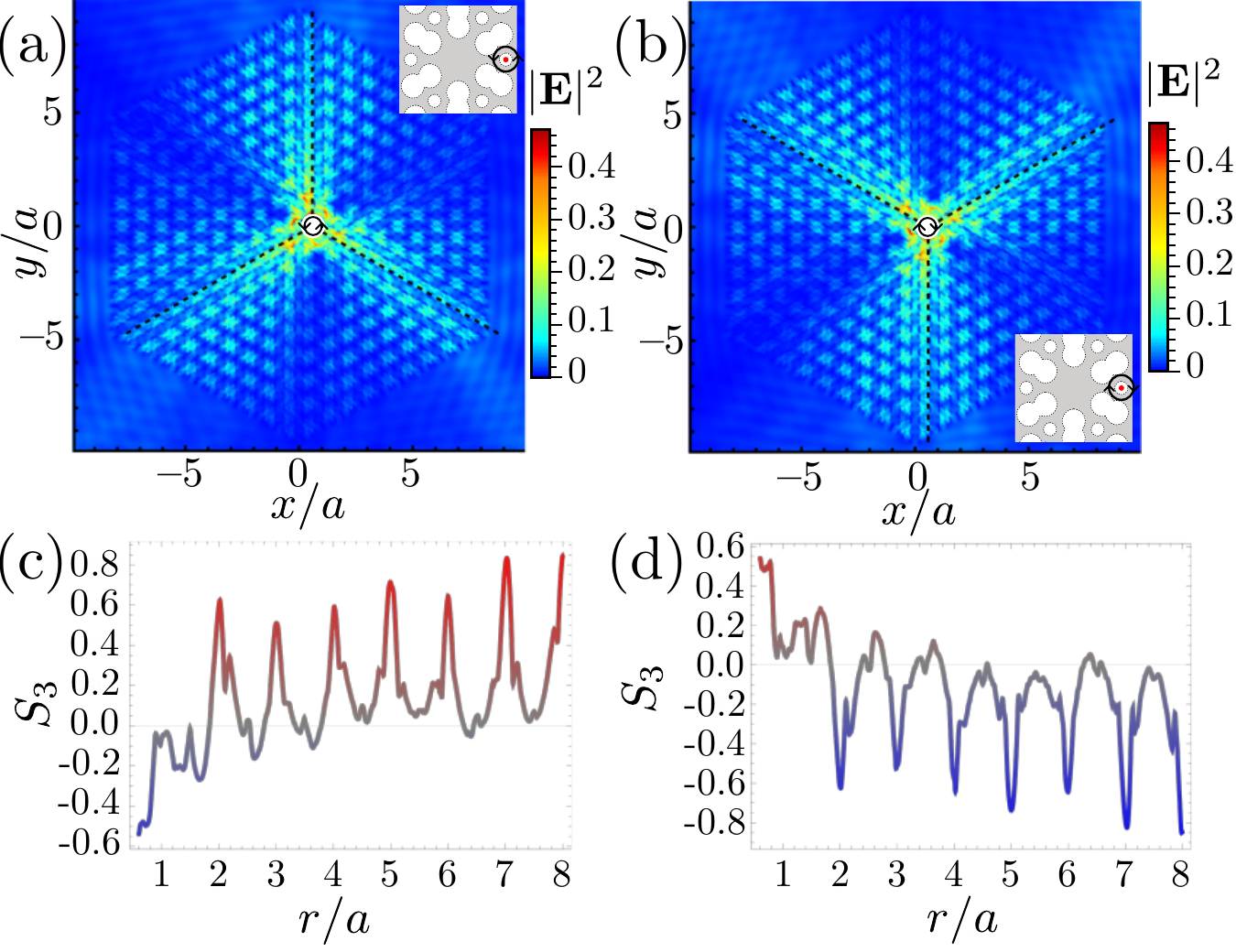}
    \caption{Circularly polarized emitter at the center of the smallest air holes. (a-b) Electric field intensity is produced for a point source with $\sigma_+$ and $\sigma_-$ dipole moment, respectively. The dashed line shows the paths that are activated in each case. The inset displays the emitter position inside the unit cell and its polarization. (c-d) Stokes parameter $S_3$ throughout the dashed line as a function of the distance from the source, for $\sigma_+$ and $\sigma_-$ dipole moment, respectively.}
    \label{fig:05circular-polar}
\end{figure}

Finally, let us analyze the case where the emitter is in the center of the smallest air hole at $\rr=(0,a/\sqrt{3})$. In contrast to the center of the unit cell, here the six principal directions are not equal; for example, if we consider the vertical direction, the photonic structure is different upward than downward from the center of air holes. We expect that this anisotropy causes an upward and downward difference in the emission patterns that can be controlled by the polarization of light, for example, considering circular polarizations $\sigma_\pm$. To test this intuition, we consider first a left-handed circular polarized ($\sigma_+$) dipole source, which selects three of the six emission paths, as shown in Fig.~\ref{fig:05circular-polar} where emitted electric field intensity is displayed. In this case, the $\sigma_+$ polarization can be used to filter the paths at $\pi/2$, $7\pi/6$, and  $11\pi/6$ rad, turning off the other main paths at $\pi/6$, $5\pi/6$ and $3\pi/2$. The choice of paths can be inverted if we consider right-handed circularly polarized ($\sigma_-$) dipoles, where the paths activated are at $\pi/6$, $5\pi/6$ and $3\pi/2$ rad, while the inhibited paths are at $\pi/2$, $7\pi/6$ and  $11\pi/6$ rad, as show the intensity profile in Fig.~\ref{fig:05circular-polar}(b). The change in the electric field pattern passing from left- to right-handed circularly polarized dipole is shown in the Supplementary Material (see Video 4).

Additionally, for these two cases, $\sigma_+$ and $\sigma_-$ dipoles, we quantify the circular polarization of the radiative modes by using the Stokes parameter $S_3$~\citep{Lang2015, Sauer2020}, defined as:
\begin{equation}
    S_3= \frac{-2 \text{Im}[E_{x}E_{y}^{*}]}{E_{x}^2+E_{y}^2}= \frac{E_{-}^{2}-E_{+}^{2}}{E_{+}^2+E_{-}^2}
\end{equation}
where $E_{x(y)}$ are the Cartesian's components of the field and $E_{+(-)}$ are the components in the circular polarization basis ($\hat \sigma_{\pm}$). The $S_3=\pm 1$ indicates left and right-handed circular polarization, while intermediate values indicate both contributions that can be elliptical or linear polarization. 
The activated paths contain several centers of the smallest air holes connected by a lattice vector, specifically by a lattice parameter in the radial direction. For the case of the $\sigma_+$ emitter, each center in the three activated paths has a dominated $\sigma_+$ polarization. In contrast, the field has an almost equal contribution of both left and right polarization for other distances throughout the path, as we can observe from the $S_3$ parameter in Fig.~\ref{fig:05circular-polar}(c). For the case of $\sigma_-$, the $S_3$ parameters are plotted in Fig.~\ref{fig:05circular-polar}(d) as a function of the distance in some of the three activated paths (the rest show qualitatively similar physics), where one can observe that at each center the $\sigma_-$ dominate over the $\sigma_+$. In contrast, both contributions are similar for other distances, suggesting a possible approximately linear polarization. This interplay between directional emission and circular polarization can be used in several applications in chiral quantum optics, as anisotropic dispersions or anisotropic coupled between emitters~\citep{Yoo2015a, Sollner2015,lodahl17a}.

\section{Conclusions}\label{sec:06}

Summing up, we study the spontaneous emission patterns near a two-dimensional photonic crystal slab featuring band structures with straight iso-frequencies and saddle-points. As expected from simplified descriptions in the literature, we are able to find regimes of highly directional emission, a requisite to observe strong super and subradiant effects in two dimensions. However, thanks to our systematic study, we uncover an interesting interplay between such directional emission patterns and the emitter's polarization and position, which has passed unnoticed in other works. In particular, when the emitter is at the center of the unit cell, we find that some of the radiative directions can be canceled by aligning linearly polarized dipoles properly. On the contrary, for out-of-center positions, we find that one can activate three out of the six propagation directions by judiciously choosing the left/right-polarized nature of the emitter's dipole. This can be considered a two-dimensional generalization of the spin-orbit coupling of light appearing in chiral quantum optical setups~\cite{lodahl17a}. Our results highlight the importance of carefully considering the electric field's structure in the study of individual and collective spontaneous emission near photonic crystals and pave the way for observing unconventional collective phenomena in such systems.

%%%%%%%%%%%%%%%%%%%%%%%%%%%%%%%%%%%%%%%%%%%%%%%%%%%%%%%%%%%%%%%%%%%%%%%%%%%%%%%
%%%%%%%%%%%%%%%%%%%%%%%%%%%%%%%%%%%%%%%%%%%%%%%%%%%%%%%%%%%%%%%%%%%%%%%%%%%%%%%
		
\section{Author statements}

\begin{funding}
The authors acknowledge support from i-COOP program from CSIC with project reference COOPA20280. EPNB thanks financial support from the ``Programa de becas de excelencia doctoral del bicentenario - MINCIENCIAS 2019". AGT acknowledges support from the Proyecto Sin\'ergico CAM 2020 Y2020/TCS-6545 (NanoQuCo-CM), from the CSIC Interdisciplinary Thematic Platform (PTI) Quantum Technologies (PTI-QTEP+), from Spanish projects PID2021-127968NB-I00 and TED2021-130552B-C22 funded by MCIN/AEI/10.13039/501100011033/FEDER UE and MCIN/AEI/10.13039/501100011033, respectively, and the support from a 2022 Leonardo Grant for Researchers and Cultural Creators, BBVA. EPNB and HVP gratefully acknowledge funding from the "Centro de Excelencia en Computación Cuántica e Inteligencia Artificial", HERMES code 52893.
\end{funding}

\begin{authorcontributions}
(\emph{mandatory}) EPNB contributed with the calculations and figures of the manuscript as well as manuscript writing, while HVP and AGT supervised the project and contributed to the writing. All authors have accepted responsibility for the entire content of this manuscript and approved its submission. The datasets generated during and/or analyzed during the current study are available from the corresponding author on reasonable request.

\end{authorcontributions}

\begin{conflictofinterest}
Authors state no conflict of interest.
\end{conflictofinterest}

\bibliographystyle{ieeetr}
\bibliography{references,referencesAlex}

\end{document}